\documentclass[journal=jacsat,manuscript=article]{achemso}

\usepackage[version=3]{mhchem} 

\usepackage{float}
\DeclareUnicodeCharacter{0301}{\'{e}}
\DeclareUnicodeCharacter{2218}{\'{e}}
\DeclareUnicodeCharacter{2082}{\'{e}}

\author{Abhisek Mishra}
\affiliation{Laboratory for Nanomagnetism and Magnetic Materials, School of Physical Sciences, National Institute of Science Education and Research (NISER), An OCC of Homi Bhabha National Institute (HBNI), Jatni, 752050, India}
\author{Kshitij Singh Rathore}
\affiliation{Laboratory for Nanomagnetism and Magnetic Materials, School of Physical Sciences, National Institute of Science Education and Research (NISER), An OCC of Homi Bhabha National Institute (HBNI), Jatni, 752050, India}

\author{Swayang Priya Mahanta}
\affiliation{Laboratory for Nanomagnetism and Magnetic Materials, School of Physical Sciences, National Institute of Science Education and Research (NISER), An OCC of Homi Bhabha National Institute (HBNI), Jatni, 752050, India}

\author{Subhankar Bedanta}
\affiliation{Laboratory for Nanomagnetism and Magnetic Materials, School of Physical Sciences, National Institute of Science Education and Research (NISER), An OCC of Homi Bhabha National Institute (HBNI), Jatni, 752050, India}
\altaffiliation{Center for Interdisciplinary Sciences (CIS), NISER, An OCC of HBNI, Jatni, 752050, India}

\email{sbedanta@niser.ac.in}

\title[An \textsf{achemso} demo]
  {Efficient spin to charge conversion and spin memory loss mitigation in oriented RuO$_2$ films}


\begin{document}

\begin{abstract}
   RuO$_2$, a transition metal oxide, is attracting attention in spintronics for its unique altermagnetic properties, which influence spin currents. Its ability to produce large spin-orbit torques and spin Hall effects is key for energy-efficient magnetic memory and logic devices. Additionally, the tunable thickness and crystallinity of RuO$_2$ thin films optimize torque efficiency for low-power switching. Spin pumping, a versatile method for investigating spin dynamics in RuO$_2$ thin films, has garnered considerable interest because of its straightforward, non-invasive and uncomplicated approach to addressing impedance mismatch and direct measurement of spintronic parameters. Here we present a systematic and detailed analysis on the efficient spin to charge conversion in (110)-oriented RuO$_2$ films with amorphous CoFeB as spin source. The spin Hall angle, and spin diffusion length were estimated to be 0.14 $\pm$ 0.01 and 4.58 $\pm$ 0.40 nm, respectively. The spin Hall conductivity of 998.89 $\pm$ 58.23 $\frac{\hbar}{e}\Omega^{-1}$cm$^{-1}$  has been estimated which is theoretically predicted to be of the similar order. The interfacial spin transparency has been achieved to be 90$\%$. We have shown that the spin memory loss at the RuO$_2$/CoFeB interface is 15$\%$, which is very small.
   \section{Keywords:} RuO$_2$, ISHE, Interfacial spin transparency, spin Hall conductivity, FMR, Thin films
\end{abstract}


\section{Introduction}
Spintronic devices and technologies harness the intrinsic spin of electrons in addition to their charge to achieve enhanced functionality and performance. A fundamental phenomenon in spintronics is spin-orbit coupling (SOC), which arises from the interaction between an electron's spin and its orbital motion in a material \cite{hirohata2020review,soumyanarayanan2016emergent,ryu2020current}. This coupling can lead to various effects that are crucial for spintronics applications. One such effect is the spin Hall effect (SHE), where an applied charge current generates a transverse spin current due to spin-orbit coupling \cite{hirsch1999spin}. Conversely, its Onsager reciprocal the inverse spin Hall Effect (ISHE) describes how a spin current can be converted into a charge current, also mediated by spin-orbit coupling \cite{saitoh2006conversion}. These effects are instrumental in spin-to-charge conversion processes, which are essential for efficient signal transduction in spintronic devices. In this context spin pumping, plays a critical role especially in generating the pure spin current which is basically the flow of spin angular momentum \cite{tserkovnyak2002spin}. Spin pumping is integral to the performance of various spintronic devices, such as spin-torque oscillators, magnetic tunnel junctions and magnetic random access memories (MRAMs). It helps in controlling and modulating spin currents, thereby affecting the device's functionality and efficiency. In spin-orbit torque-based MRAMs, a charge current in the nonmagnetic metal layer produces a spin current directed orthogonally, which is then utilized to precisely control the magnetization in the adjacent ferromagnetic layer \cite{shao2021roadmap}. Here, the goal is to find a suitable material which can help generate and manipulate the pure spin currents. Heavy metals along with various quantum materials such as topological insulators, transition metal dichalcogenides, antiferromagnets, etc have been reported to show substantial SOC \cite{roy2021spin,singh2020inverse,singh2020large,kimata2019magnetic,singh2019inverse,jamali2015giant}.
TMOs, in this context, provide pronounced SOC strength \cite{chen2021spin}. The family of transition metal oxides possesses a range of properties that make them highly attractive as spin-Hall materials. Spin-Hall materials typically require a high spin-Hall angle and low or moderate resistivity. TMOs such as IrO$_2$ possess a relatively large spin-Hall angle combined with intermediate resistivity, making them promising candidates to bridge the gap between heavy metals and topological materials \cite{sahoo2021spin}. Moreover, these oxides exhibit strong coupling between lattice, orbital, spin, and charge degrees of freedom \cite{tokura2000orbital}. It is well established that the crystal symmetry of transition metal oxides can be tuned through various methods which is not available in other materials, opening new pathways for engineering spin–charge conversion \cite{medvedeva2010tuning}. The properties of TMOs can be tuned through growth processes, doping, strain, and structural modifications, allowing for precise control over their electronic and magnetic characteristics \cite{li2020defects,cao2011strain}. The discovery of effective spin-Hall materials within this family could pave the way for integrating spin–charge conversion with other functionalities by forming high-quality heterostructures. 
Recently, RuO$_2$ has achieved significant scientific interest in spintronics owing to its altermagnetic ordering, high SOC and spin Hall conductivity \cite{vsmejkal2022emerging}. For instance, epitaxial RuO$_2$ films of different crystallographic orientations have shown substantial anomalous Hall conductivity \cite{feng2022anomalous}. Zhang et al showed that (101) oriented RuO$_2$ films show large spin Hall angle and long spin diffusion length \cite{zhang2024simultaneous}. Bose et al, proved that symmetry plays a central role in determining the polarization of spin currents induced by electric fields and hence showed the generation of tilted spin current in (101) oriented RuO$_2$ which generated an out-of-plane damping like torque \cite{bose2022tilted}. Bai et al demonstrated a current induced spin splitting effect in RuO$_2$ films where the direction of spin current is correlated to the crystallographic orientation of RuO$_2$ \cite{bai2022observation}. This implies that the different orientations of RuO$_2$ show variation of numerous physical properties. 
Despite recent progress, there is a significant lacking of both qualitative and quantitative analysis on spin-to-charge conversion and the quantification of interfacial spintronic parameters in (110)-oriented RuO$_2$ film-based heterostructures. In this study, we report a systematic analysis of spin to charge conversion via spin pumping. Additionally, we conducted a detailed study of thickness and angle-dependent ISHE measurements in these bilayers to separate spin rectification effects from spin pumping voltage. The RuO$_2$ films have been prepared by reactive magnetron sputtering of a Ru target in the presence of O$_2$ instead of a stoichiometric RuO$_2$ target. This highlights a more controlled and flexible method of film deposition. Reactive magnetron sputtering using a Ru target with O$_2$ allows for fine-tuning the oxygen content during the deposition process, which can lead to better control over the film's stoichiometry, quality, and properties. Further, crucial spintronic parameters such as, spin pumping voltage, spin Hall angle, spin Hall conductivity, spin diffusion length, spin mixing conductance, interfacial spin transparency and spin memory loss have been estimated.

\section{Experimental details}
Thin films were prepared using a high-vacuum multi-deposition chamber (Excel Instruments, India) with a base pressure below 5$\times$10$^{-8}$ mbar. RuO$_2$ films were deposited onto MgO (100) substrates by sputtering a Ru target in the presence of a gas mixture of 30 sccm oxygen and 20 sccm argon. During deposition, the substrate table was rotated at 20 rpm to ensure uniformity, and the growth temperature was maintained at 500 $^\circ$C. Post-deposition, the films were annealed at the same temperature of 500 $^\circ$C for 5 minutes in an oxygen environment. The chamber was then allowed to evacuate the remaining oxygen and cool the substrate to room temperature before the deposition of metallic ferromagnet.
Subsequently, Co$_{40}$Fe$_{40}$B$_{20}$ (CoFeB) film was deposited using a commercially available stoichiometric target. To prevent oxidation, a layer of AlO$_X$ was deposited on CoFeB via \textit{rf} magnetron sputtering. The resulting bilayer samples were composed of RuO$_2$ ($t_{RuO_2}$)/CoFeB (9 nm)/AlO$_X$(3 nm), with the RuO$_2$ thickness ($t_{RuO_2}$) varying between 0 and 20 nm. These samples were labeled as S0, S1, S2, S3, S4, S5, and S6, corresponding to RuO$_2$ thicknesses of 0, 2, 3, 5, 7, 10, and 20 nm, respectively as shown in Table 1.
X-ray diffractometry of the samples was performed with a Rigaku diffractometer (Fig. 1 (b)). Figure 1 (a) shows the crystal structure of RuO$_2$. For ferromagnetic resonance (FMR) experiments, the samples were positioned in a flip-chip orientation on a coplanar waveguide (CPW) (Fig. S1). An in-plane DC magnetic field (\textit{H}) was applied perpendicular to the \textit{rf} field ($h_{rf}$) using an electromagnet. The Gilbert damping constant ($\alpha$) was determined by measuring FMR spectra at frequencies ranging from 4 to 17 GHz in 0.5 GHz intervals. The DC voltage ($V_{dc}$) generated by the inverse spin Hall effect (ISHE) was recorded using a nanovoltmeter integrated into a custom-modified FMR setup \cite{singh2019inverse}. All measurements were conducted on samples approximately 2 $\times$ 3 mm$^2$ in size, with electrical contacts made at the longer edges using silver paste and copper wires. Angle-dependent ISHE measurements were performed at a microwave frequency of 7 GHz using a motorized rotation stage. Microwave power-dependent ISHE measurements were conducted using an SMB-100 \textit{rf} signal generator from RHODE $\&$ SCHWARZ. The saturation magnetization was measured using a SQUID-based magnetometer (MPMS 3, Quantum Design).
\begin{table}[]
\caption{Samples studied for the present work (The numbers in the bracket are in the units of nm).}
        \centering
\begin{tabular}{|c|c|}
\hline

S0 & CoFeB(9)/Al$_2$O$_3$(3)                                                     \\ \hline
S1 & RuO$_2$(2)/CoFeB(9)/Al$_2$O$_3$(3)          \\ \hline
S2 & RuO$_2$(3)/CoFeB(9)/Al$_2$O$_3$(3)  \\ \hline
S3 & RuO$_2$(5)/CoFeB(9)/Al$_2$O$_3$(3) \\ \hline
S4 & RuO$_2$(7)/CoFeB(9)/Al$_2$O$_3$(3)    \\ \hline
S5 & RuO$_2$(10)/CoFeB(9)/Al$_2$O$_3$(3)  \\ \hline
S6 & RuO$_2$(20)/CoFeB(9)/Al$_2$O$_3$(3) \\ \hline
\end{tabular}
	\label{table1}
\end{table}

\begin{figure}[ht]
	\centering
	\includegraphics[width=0.9\textwidth]{"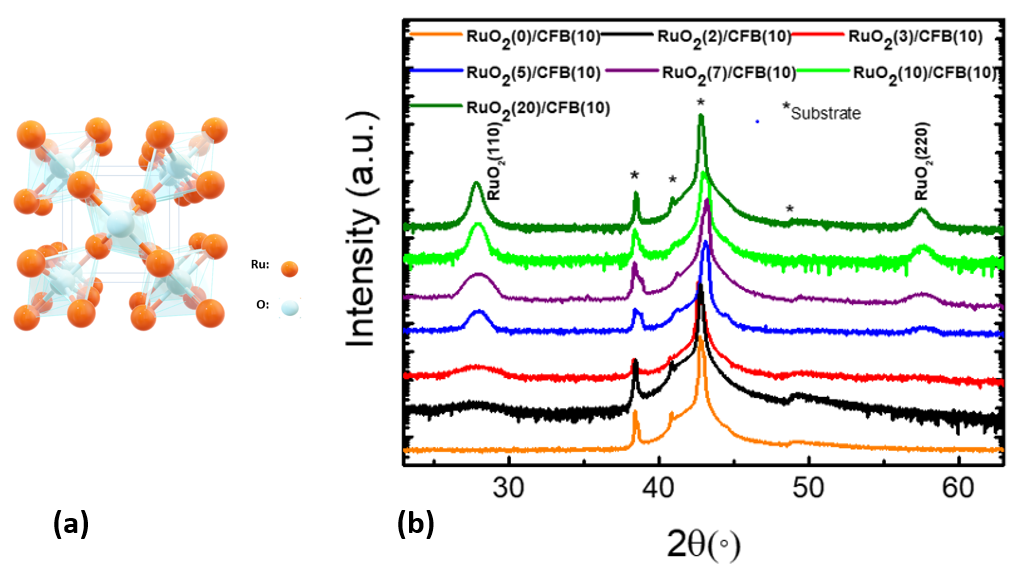"}
	\caption{(a) Crystal structure of RuO$_2$ (b) X-ray diffraction peaks of RuO$_2$/CFB samples}
	\label{fig:figure-1}
\end{figure}

\section{Results and discussion}
Fig. 1 (b) shows the X-ray diffraction patterns of RuO$_2$/CoFeB samples on MgO (100) substrates. Since CoFeB is amorphous in nature, it did not show any peaks. However, we observe diffraction peaks of RuO$_2$ corresponding to (110) planes. Using the FMR measurements shown in Fig. S2 (see supplementary information), the resonance field ($H_{res}$) and linewidth (\textbf{$\Delta H$}) were obtained by fitting the frequency-dependent experimental data to a Lorentzian derivative function. The gyromagnetic ratio ($\gamma$) was then calculated by fitting the \textit{f} vs. $H_{res}$ data, as illustrated in Fig. 2(a), using Kittel's equation, which is expressed as follows \cite{kittel1948theory},
\begin{equation}\label{equation1}
    \it{f}=\frac{\gamma}{2 \pi} \sqrt{(H_K+H_{res})(H_K +H_{res} + 4 \pi M_{eff})}
\end{equation}
Here, $H_K$ and $4{\pi{M_{eff}}}$ denote the anisotropy field and effective magnetization, respectively. The Gilbert damping constant $\alpha$ was determined by fitting the data in Fig. 2(b) using the following linear equation:

\begin{equation}\label{equation2}
    \Delta H = \Delta H_0 +\frac{4 \pi \alpha \it{f}}{\gamma}
\end{equation}
\begin{figure}[ht]
	\centering
	\includegraphics[width=0.9\textwidth]{"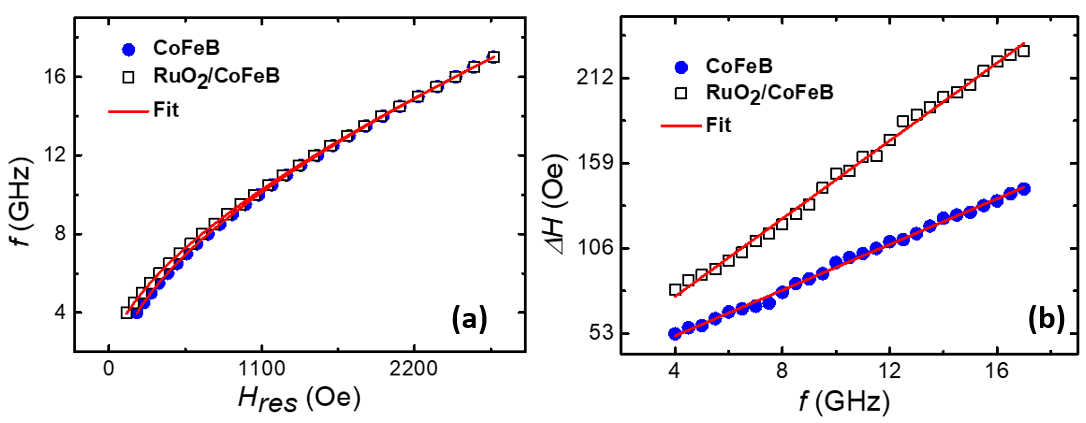"}
	\caption{(a) \textit{f} versus $H_{res}$. (b) \textbf{$\Delta H$} versus \textit{f} for the samples S3 (squares) and S0 (solid circles). The solid red lines in (a) and (b) are best fits to eqs. (1) and (2), respectively.}
	\label{fig:figure-2}
\end{figure}

Here, \textbf{$\Delta H_0$} represents the y-intercept of the linear equation, indicating the extent of magnetic inhomogeneity within the samples. The effective $\alpha$ values are also impacted by several factors, including spin pumping, magnetic proximity effects (MPEs), interface phenomena, impurities, and other factors that can contribute to an increase in the system's damping constant \cite{brataas2002spin}. The $\alpha$ ($\times$10$^{-3}$) values of the samples S1-S6 are 14.12$\pm$ 0.13, 15.30$\pm$ 0.24, 17.61$\pm$ 0.14, 16.31$\pm$ 0.23, 17.82$\pm$ 0.21, and 18.21$\pm$ 0.21,respectively, which are higher than that of the reference layer S0 (8.71$\pm$ 0.16). The increased damping values observed in the bilayer samples indicate possible presence of  spin pumping. However, the influence of other effects cannot be ruled out. To verify the presence of spin pumping, we performed ISHE measurements on these samples. Fig. 2(a) shows how the measured voltage due to ISHE varies with the applied field H for the S3 sample at 0 and 180 degrees. The symmetric ($V_{sym}$) and antisymmetric ($V_{asym}$) components of the voltage signal ($V_{dc}$) were isolated by fitting the data to the Lorentzian equation given below \cite{conca2017lack},
\begin{equation}
\begin{aligned}
    V_{dc} = V_{sym} \frac{(\Delta H)^2}{(H-H_{res})^2+(\Delta H)^2}+\\V_{asym} \frac{2 \Delta H (H - H_{res})}{(H-H_{res})^2+(\Delta H)^2}
    \end{aligned}
\end{equation}

\begin{figure}[ht]
	\centering
	\includegraphics[width=1\textwidth]{"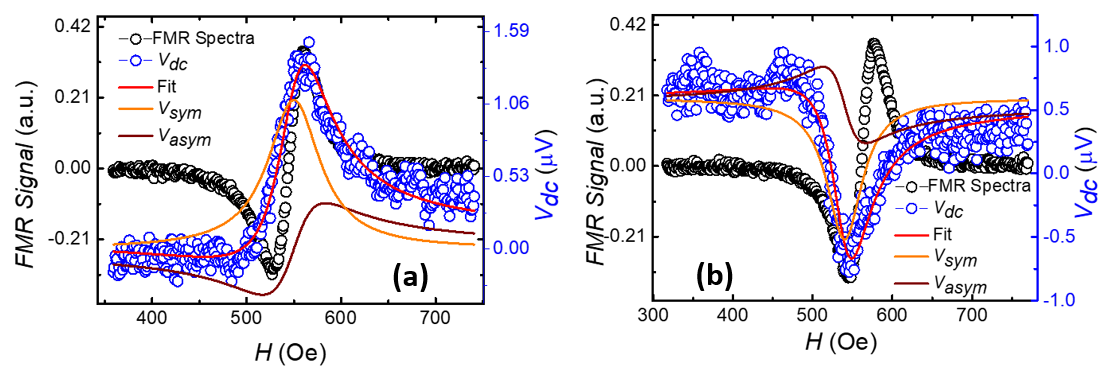"}
	\caption{Voltage, $V_{dc}$ (open blue symbols) measured across the sample with applied magnetic field along with FMR signal (open black symbols) for sample S3 at the $\phi$ values of (a) 0$^\circ$ and (b) 180$^\circ$. Solid red lines are the fit to the experimental data using equation (3). Solid orange and brown lines are the symmetric and anti-symmetric components of the voltage.}
 \label{fig:figure-3}
\end{figure}
Fig. 4(a) and (b) depict the measured voltage as a function of the applied DC field \textit{H} for sample S3, with data recorded at in-plane angles of $\phi$ = 0$^\circ$ and 180$^\circ$, respectively. In this context, $\phi$ represents the angle between the measured voltage direction and the direction perpendicular to the applied field H. The reversal in the sign of $V_{dc}$ as $\phi$ changes from 0$^\circ$ to 180$^\circ$ clearly indicates the presence of spin pumping in our samples. In spin pumping experiments, where the goal is to study the generation and detection of spin currents and their conversion into charge currents by ISHE, spin rectification effects (SRE) can introduce spurious signals. These signals may be misinterpreted as contributions from spin pumping, leading to incorrect conclusions about the efficiency or nature of spin current generation. Spin rectification effects occur when the interaction between an oscillating magnetization and an external radio frequency field generates a \textit{dc} voltage in magnetic materials. These effects, such as those arising from anisotropic magnetoresistance (AMR) or the anomalous Hall effect (AHE), can produce signals that mimic or interfere with those generated by spin pumping. To accurately interpret the results of spin pumping experiments, it is crucial to distinguish between genuine spin pumping-induced signals and those arising from spin rectification effects.
To separate the unwanted SREs, angle-dependent ISHE measurements were performed to isolate the contributions of AHE and AMR. Fig. 3(a) and (b) display the angle-dependent Vsym  and Vasym  components, respectively, for sample S3. These plots were fitted using the equations given below \cite{conca2017lack},
\begin{equation}
\begin{aligned}
{V_{sym}= V_{sp}cos^3(\phi + \phi_0)+V_{AHE}\ cos(\theta)}cos(\phi + \phi_0)\\
    + V_{sym}^{AMR \perp} cos 2(\phi + \phi_0)cos(\phi+ \phi_0)\\
    + V_{sym}^{AMR ||}sin2(\phi + \phi_0)cos(\phi+\phi_0)
    \end{aligned}
\end{equation}
\begin{equation}
\begin{aligned}
 V_{asym}= V_{AHE}\ sin (\theta) cos(\phi + \phi_0) + 
    \\V_{asym}^{AMR \perp} cos 2(\phi + \phi_0)cos(\phi + \phi_0)+ 
   \\ V_{asym}^{AMR ||}sin2(\phi + \phi_0)cos(\phi+\phi_0)
   \end{aligned}
 \end{equation}
 \begin{figure}[ht]
	\centering
	\includegraphics[width=0.9\textwidth]{"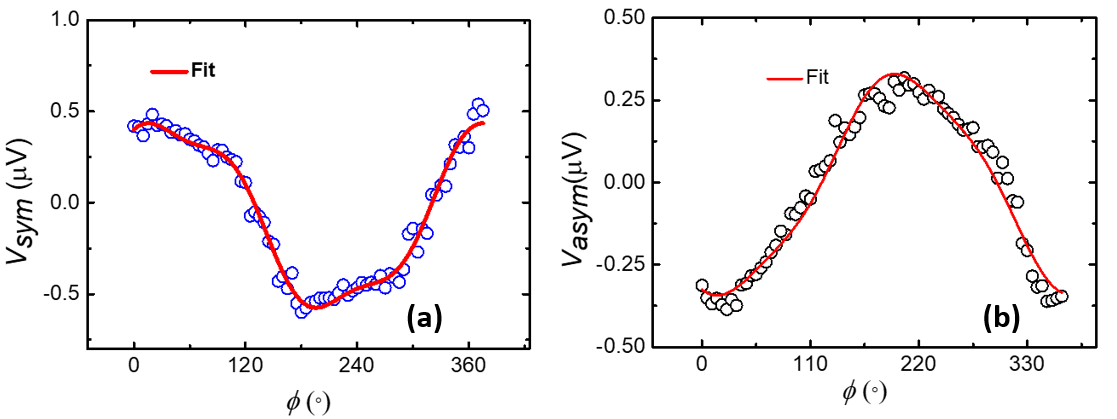"}
	\caption{(a) $\phi$ dependent (a) $V_{sym}$ and (b) $V_{asym}$ for sample S4. Figure (a) is fitted to Eq. (4), while Figure (b) is fitted to Eq. (5).}
	\label{fig:figure-4}
\end{figure}
An additional factor, $\phi_0$, was introduced to account for any misalignment in sample positioning during the measurement of $\phi$ values. In this context, $\theta$ represents the angle between the electric and magnetic fields of the applied microwave, which is 90$^\circ$. The parameters for the various components, as determined from the angle-dependent ISHE measurements, are summarized in Table I.
\begin{table*}[t]
\caption{Fitted parameters from the in-plane angle dependent ISHE measurements}

\centering
\begin{tabular}{ccccc} \label{table2}

Sample & $V_{sp}$(V)$\times$$10^{-6}$ & $V_{AHE}$(V)$\times$$10^{-6}$ & $V_{AMR}^{\perp}$(V)$\times$$10^{-6}$ & $V_{AMR}^{||}$(V)$\times$$10^{-6}$ \\ \hline
M1     & 2.98 $\pm$ 0.08              & 0.18 $\pm$ 0.06              & 2.31 $\pm$ 0.08                       & 0.012 $\pm$ 0.001                    \\ \hline
M2     & 0.97 $\pm$ 0.02              & 0.32 $\pm$ 0.01              & 0.54 $\pm$ 0.03                       & 0.002 $\pm$ 0.001                    \\ \hline
M3     & 1.68 $\pm$ 0.05              & 0.16 $\pm$ 0.02              & 1.08 $\pm$ 0.06                       & 0.13 $\pm$ 0.03                    \\ \hline
M4     & 2.16 $\pm$ 0.11              & 0.48 $\pm$ 0.04              & 1.53 $\pm$ 0.12                       & 0.11 $\pm$ 0.05                    \\ \hline
M5     & 5.70 $\pm$ 0.12              & 0.49 $\pm$ 0.01              & 3.36 $\pm$ 0.13                       & 0.17 $\pm$ 0.06                    \\ \hline
M6     & 4.45 $\pm$ 0.13              & 1.72 $\pm$ 0.05              & 2.63 $\pm$ 0.13                       & 0.28 $\pm$ 0.06                    \\ \hline
M7     & 4.28 $\pm$ 0.05              & 1.06 $\pm$ 0.04              & 2.77 $\pm$ 0.05                       & 0.11 $\pm$ 0.02                    \\ 
\end{tabular}

\end{table*}
It has been observed that $V_{sp}$ significantly outweighs rectification effects, indicating robust spin pumping. To further verify that the detected voltage is primarily due to spin pumping, microwave power-dependent measurements were conducted up to $\sim$ 160 mW (see Fig. S4 (a) in supplementary information). As the microwave power P increases, the amplitude of the \textit{rf} microwave field rises proportionally to $\sqrt{P}$ \cite{mosendz2010detection} . The magnetization precession cone angle increases linearly with the \textit{rf} field, and since the voltage from the ISHE is quadratic with respect to the cone angle, the voltage is expected to increase with power. We calculated the precession cone angle of magnetization, given by $\theta_P$ = $h_{rf}$/$\Delta H$. The values of $\theta_P$ range from 0.049$\pm$0.001 to 0.111$\pm$0.002 was the microwave power increases from 31 to 160 mW. The resulting voltage displayed a $\sin^2(\theta_P)$ dependence (fig. S4 (b)) \cite{vlaminck2013dependence}. Our experiments are performed in the linear excitation regime and it shows the absence of non-uniform modes. 
The effective spin mixing conductance, which governs the magnitude of spin current that can traverse the interface, was calculated using the following expression:
\begin{equation}
 g_{eff}^{\uparrow\downarrow}=\frac{\Delta\alpha 4\pi M_{s}t_{CoFeB}}{g\mu_{B}} 
\end{equation}
where \textit{$\Delta\alpha$}, \textit{$M_S$}, \textit{$t_{CoFeB}$}, \textit{$\mu_{B}$}, and \textit{g} represent the change in $\alpha$ compared to the reference CoFeB layer, the saturation magnetization obtained from SQUID magnetometry, the thickness of the CoFeB layer, the Bohr magneton, and the Landé \textit{g}-factor, respectively.
$g_{eff}^{\uparrow\downarrow}$ reflects how well spin information is preserved across the interface, which is crucial for designing efficient spintronic devices. The measured  $g_{eff}^{\uparrow\downarrow}$($\times$ 10$^{19}$m$^{-2}$) values for the samples S1-S6 are 2.45$\pm$0.12, 2.80$\pm$0.08, 27.10$\pm$0.10, 3.80$\pm$0.04, 26.20$\pm$0.10, 3.34$\pm$0.06, 3.99$\pm$0.07, and 4.18$\pm$0.05 , respectively. 
The intrinsic spin-mixing conductance $g^{\uparrow\downarrow}$ is calculated from the fitting of the RuO$_2$ thickness dependent damping enhancement using the equation (Fig. 5)\cite{shaw2012determination,foros2005scattering},
\begin{equation}
 \Delta\alpha= \alpha_{eff}-\alpha_{CoFeB}=g\frac{g^{\uparrow\downarrow}\mu_{B}}{4\pi{M_S}{t_{CoFeB}}}[1-e^{\frac{-2t_{MoS_2}}{\lambda_{SD}}}]
\end{equation}
where $\alpha_{CoFeB}$ is the damping constant of the reference sample S0, $t_{RuO_2}$ is the thickness of RuO$_2$ in the bilayer samples and $\lambda_{SD}$ is the spin diffusion length of RuO$_2$. Here, $g^{\uparrow\downarrow}$ represents the intrinsic spin mixing conductance, which does not take spin backflow into consideration. In contrast, $g_{eff}^{\uparrow\downarrow}$ includes the effects of spin angular momentum coming back from the non-magnetic layer to the ferromagnetic layer at both interfaces.
\begin{figure}[ht]
	\centering
	\includegraphics[width=0.65\textwidth]{"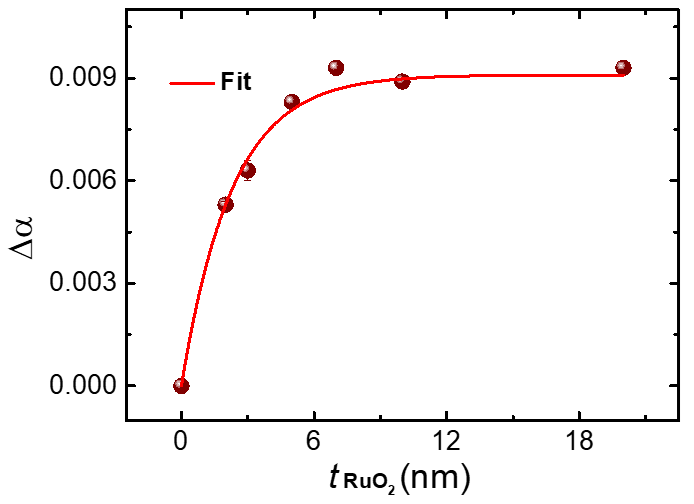"}
	\caption{Enhancement in the damping constant vs. RuO$_2$ layer thickness. Solid line is a fit to equation (7)}
	\label{fig:figure-5}
\end{figure}
From the fitting, the $\lambda_{SD}$ and $g^{\uparrow\downarrow}$ are found to be 4.58 $\pm$ 0.40 nm and 4.99 $\pm$ 0.04 $\times$ 10$^{19}$m$^{-2}$, respectively. We have calculated $\theta_{SHA}$ by using the below equation,
\begin{equation}
    \frac{V_{SP}}{R} = w\times \theta_{SHA} \lambda_{SD}tanh(\frac{t_{MoS_2}}{2 \lambda_{SD}}) {J}_s
\end{equation}
where \textit{$J_S$} is given by,
\begin{equation}
\begin{aligned}
    {J}_s \approx\frac{{g_{eff}^{\uparrow\downarrow}}{\gamma}^2 {h_{rf}}^2\hbar[4\pi M_s\gamma+\sqrt{(4\pi M_s\gamma)^2+16(\pi f)^2}]}{8\pi{\alpha}^2[{(4\pi M_s\gamma)^2+16(\pi f)^2}]}
    \end{aligned}
\end{equation}
Here, \textit{w} denotes the width of the transmission line in the coplanar waveguide (CPW) and \textit{R} is the resistance of the samples measured using the four-probe technique. The resistance of the bilayer samples S1-S6 are 245.28 $\pm$ 0.71, 220.22 $\pm$ 0.62, 160.49 $\pm$ 0.56, 140.33 $\pm$ 0.30, 91.46 $\pm$ 0.12 and 46.64 $\pm$ 0.09 $\Omega$, respectively. Since the conductivity of RuO$_2$ is higher than that of CoFeB, majority of current flows through the RuO$_2$ layer. Considering the standard Boltzmann equation we have calculated the scattering parameters and mean free path encountered by electrons in our samples (see supplementary). The Fermi wave vector ($k_f$), Fermi velocity ($k_v$) and carrier concentration (\textit{n}) were calculated to be 3.09 $\pm$ 0.29$\times$10$^7$cm$^{-1}$, 3.50 $\pm$ 0.29 $\times$ 105 m/s and 7.14 $\pm$ 0.26 $\times$ 10$^{20}$/cm$^3$, respectively which indicates the nature of a highly conducting oxide \cite{lee2011progress}. In our setup, the \textit{rf} field ({$\mu_0$}$h_{rf}$) is 0.5 Oe (at +15 dBm \textit{rf} power), and the transmission linewidth, \textit{w}  is 200 $\mu$m. The estimated values of $\theta_{SHA}$ for the samples S1-S6 are 0.10, 0.12, 0.14, 0.09, 0.08 and 0.05, respectively. The spin Hall conductivity was calculated using the expression $\sigma_{SH}$ = $\sigma$ $\times$ $\frac{\hbar}{e}$ $\times$ $\theta_{SHA}$, where $\sigma$ is the conductivity of the bilayer samples. $\sigma_{SH}$ in our samples varies from 556 to 936 $\frac{\hbar}{e}\Omega^{-1}$cm$^{-1}$, which is of the similar order as calculated theoretically \cite{bose2022tilted}. 
Further, to calculate the spin flip probability and spin memory loss, we followed the below model \cite{rojas2014spin},
\begin{equation}
g_{eff}^{\uparrow\downarrow} = g_{r}^{\uparrow\downarrow} \frac{r_{sI} \cosh(\delta) + r_{sN}^{\infty} \coth \left(\frac{t_{RuO_2}}{\lambda_{SD}}\right) \sinh(\delta)}{r_{sI} \left(1 + 0.5 {\sqrt{\frac{3}{\epsilon}}} \coth \left(\frac{t_{RuO_2}}{\lambda_{SD}}\right)\right) \cosh(\delta) + \left(r_{sN}^{\infty} \coth \left(\frac{t_{RuO_2}}{\lambda_{SD}}\right) + 0.5 \frac{r_{sI}^2}{r_{sN}^{\infty}}\sqrt{\frac{3}{\epsilon}}\right) \sinh(\delta)} 
\end{equation}
where $\epsilon$ is the ratio of the spin conserved to spin-flip relaxation times and $g_{r}^{\uparrow\downarrow}$ is the real part of spin mixing conductance. In addition, $r_{sI}$, $r_{sN}^{\infty}$, and $\delta$ are the interfacial spin resistance, RuO$_2$ spin resistance and spin-flip parameter for the RuO$_2$/CoFeB interface, respectively.
\begin{figure}[ht]
	\centering
	\includegraphics[width=0.65\textwidth]{"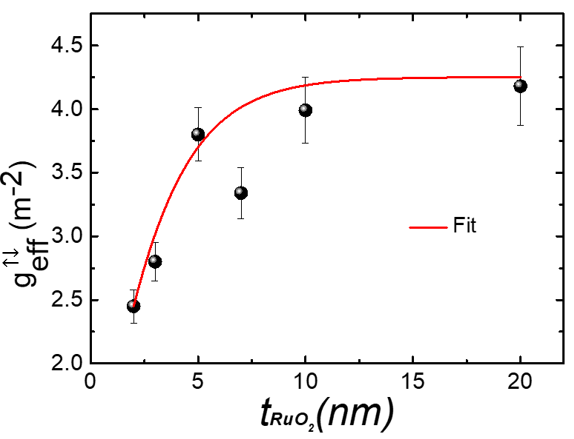"}
	\caption{$g_{eff}^{\uparrow\downarrow}$ as a function of thickness of RuO$_2$. The solid red line is the fit to equation 11}
	\label{fig:figure-6}
\end{figure}
Figure 6 shows the variation of $g_{eff}^{\uparrow\downarrow}$ as a function of RuO$_2$ thickness. The values of $g_{r}^{\uparrow\downarrow}$,$r_{sI}$,$r_{sN}^{\infty}$, $\delta$ and $\epsilon$ were found to be 3.11$\times$10$^{19}$m$^{-2}$, 5.8 f$\Omega$m$^{-2}$, 4.8 f$\Omega$m$^{-2}$,0.17 and 0.021, respectively. The spin memory loss factor is given by 1-e$^{-\delta}$ which at the interface is only 15$\%$. This is comparable to the Co/Ru system \cite{khasawneh2011spin}. Majority of the spins get transferred to the bulk of RuO$_2$ instead of getting lost at the interface. This suggests that the disorder at the interfaces is minimal, as spin depolarization is primarily due to interface disorder \cite{singh2021high}. The $\epsilon$ parameter of 0.021 indicates an efficient spin sink behaviour of RuO$_2$ as a better spin sink has $\epsilon$ $\geq$ 0.01 \cite{tserkovnyak2002spin,panda2021structural}. This is crucial for spintronic devices, as it ensures that the spin currents generated in one material can be effectively used in another without significant degradation. 
The interfacial spin transparency ($T_{int}$) is calculated using the expression \cite{zhang2015role},

\begin{equation}
T_{\text{int}} = \frac{g_r^{\uparrow\downarrow} \tanh\left(\frac{t_{\text{RuO}_2}}{2\lambda_{SD}}\right)}{g_r^{\uparrow\downarrow} \coth\left(\frac{t_{\text{RuO}_2}}{\lambda_{SD}}\right) + \frac{h}{2\rho \lambda_{SD} e^2}}
\end{equation}
where $\rho$ is the resistivity of RuO$_2$, \textit{h} is the Planck’s constant and \textit{e} is the charge of electron.  The resistivity of RuO$_2$ films of thicknesses 2, 3 ,5 ,7 ,10 and 20 nm are 273, 254, 186, 174, 132 and 111 $\mu$cm, respectively. The maximum $T_{int}$ value was found to be $\sim$ 90$\%$ for the sample S6, which indicates minimal spin relaxation or dephasing at the interface. High interfacial spin transparency contributes to better performance metrics like higher magnetoresistance ratios, faster switching times, and lower energy consumption. It ensures that the spin-polarized current remains coherent as it passes through different layers of the device.

\section{Conclusion}

We have synthesized (110)-oriented RuO$_2$ films using reactive magnetron sputtering. This study underscores the significant potential of RuO$_2$ thin films in advancing spintronic technologies. As promising candidates for spin-to-charge conversion in spintronics, \textit{4d} transition metal oxides (TMOs) demonstrate considerable utility. The achievements observed in these simple binary oxides emphasize the potential of \textit{4d} TMOs as a novel class of spintronic materials. The strong spin-orbit coupling (SOC) inherent in this material class leads to a wide range of intriguing physical properties. Through systematic analysis, we have precisely calculated key spintronic parameters, which are crucial for developing next-generation low-power magnetic memory and logic devices. The findings from spin pumping experiments reveal that our RuO$_2$-based heterostructures exhibit minimal spin memory loss and high interfacial spin transparency while facilitating efficient spin-to-charge conversion.
\section{Supporting Information}
XRD measurements, FMR spectra, power dependent measurements, resistivity analysis
\section{Author Information}
\subsection{Corresponding Author}
\textbf{Subhankar Bedanta}- Laboratory for Nanomagnetism and Magnetic Materials, School of Physical Sciences, National Institute of Science Education and Research (NISER), An OCC of Homi Bhabha National Institute (HBNI), Jatni, 752050, India

Email: sbedanta@niser.ac.in

\subsection{Authors}
\textbf{Abhisek Mishra}- Laboratory for Nanomagnetism and Magnetic Materials, School of Physical Sciences, National Institute of Science Education and Research (NISER), An OCC of Homi Bhabha National Institute (HBNI), Jatni, 752050, India

Email: abhisek.mishra@niser.ac.in

\textbf{Swayang Priya Mahanta}- Laboratory for Nanomagnetism and Magnetic Materials, School of Physical Sciences, National Institute of Science Education and Research (NISER), An OCC of Homi Bhabha National Institute (HBNI), Jatni, 752050, India

Email: swayangpriya.mahanta@niser.ac.in

\textbf{Kshitij Singh Rathore}- Laboratory for Nanomagnetism and Magnetic Materials, School of Physical Sciences, National Institute of Science Education and Research (NISER), An OCC of Homi Bhabha National Institute (HBNI), Jatni, 752050, India

Email: kshitij.rathore@niser.ac.in

\begin{acknowledgement}

We acknowledge the financial support by the Department of Atomic Energy (DAE).

\end{acknowledgement}

\bibliography{references}

\providecommand{\latin}[1]{#1}
\makeatletter
\providecommand{\doi}
  {\begingroup\let\do\@makeother\dospecials
  \catcode`\{=1 \catcode`\}=2 \doi@aux}
\providecommand{\doi@aux}[1]{\endgroup\texttt{#1}}
\makeatother
\providecommand*\mcitethebibliography{\thebibliography}
\csname @ifundefined\endcsname{endmcitethebibliography}  {\let\endmcitethebibliography\endthebibliography}{}
\begin{mcitethebibliography}{38}
\providecommand*\natexlab[1]{#1}
\providecommand*\mciteSetBstSublistMode[1]{}
\providecommand*\mciteSetBstMaxWidthForm[2]{}
\providecommand*\mciteBstWouldAddEndPuncttrue
  {\def\EndOfBibitem{\unskip.}}
\providecommand*\mciteBstWouldAddEndPunctfalse
  {\let\EndOfBibitem\relax}
\providecommand*\mciteSetBstMidEndSepPunct[3]{}
\providecommand*\mciteSetBstSublistLabelBeginEnd[3]{}
\providecommand*\EndOfBibitem{}
\mciteSetBstSublistMode{f}
\mciteSetBstMaxWidthForm{subitem}{(\alph{mcitesubitemcount})}
\mciteSetBstSublistLabelBeginEnd
  {\mcitemaxwidthsubitemform\space}
  {\relax}
  {\relax}

\bibitem[Hirohata \latin{et~al.}(2020)Hirohata, Yamada, Nakatani, Prejbeanu, Di{\'e}ny, Pirro, and Hillebrands]{hirohata2020review}
Hirohata,~A.; Yamada,~K.; Nakatani,~Y.; Prejbeanu,~I.-L.; Di{\'e}ny,~B.; Pirro,~P.; Hillebrands,~B. Review on spintronics: Principles and device applications. \emph{Journal of Magnetism and Magnetic Materials} \textbf{2020}, \emph{509}, 166711\relax
\mciteBstWouldAddEndPuncttrue
\mciteSetBstMidEndSepPunct{\mcitedefaultmidpunct}
{\mcitedefaultendpunct}{\mcitedefaultseppunct}\relax
\EndOfBibitem
\bibitem[Soumyanarayanan \latin{et~al.}(2016)Soumyanarayanan, Reyren, Fert, and Panagopoulos]{soumyanarayanan2016emergent}
Soumyanarayanan,~A.; Reyren,~N.; Fert,~A.; Panagopoulos,~C. Emergent phenomena induced by spin--orbit coupling at surfaces and interfaces. \emph{Nature} \textbf{2016}, \emph{539}, 509--517\relax
\mciteBstWouldAddEndPuncttrue
\mciteSetBstMidEndSepPunct{\mcitedefaultmidpunct}
{\mcitedefaultendpunct}{\mcitedefaultseppunct}\relax
\EndOfBibitem
\bibitem[Ryu \latin{et~al.}(2020)Ryu, Lee, Lee, and Park]{ryu2020current}
Ryu,~J.; Lee,~S.; Lee,~K.-J.; Park,~B.-G. Current-induced spin--orbit torques for spintronic applications. \emph{Advanced Materials} \textbf{2020}, \emph{32}, 1907148\relax
\mciteBstWouldAddEndPuncttrue
\mciteSetBstMidEndSepPunct{\mcitedefaultmidpunct}
{\mcitedefaultendpunct}{\mcitedefaultseppunct}\relax
\EndOfBibitem
\bibitem[Hirsch(1999)]{hirsch1999spin}
Hirsch,~J. Spin hall effect. \emph{Physical review letters} \textbf{1999}, \emph{83}, 1834\relax
\mciteBstWouldAddEndPuncttrue
\mciteSetBstMidEndSepPunct{\mcitedefaultmidpunct}
{\mcitedefaultendpunct}{\mcitedefaultseppunct}\relax
\EndOfBibitem
\bibitem[Saitoh \latin{et~al.}(2006)Saitoh, Ueda, Miyajima, and Tatara]{saitoh2006conversion}
Saitoh,~E.; Ueda,~M.; Miyajima,~H.; Tatara,~G. Conversion of spin current into charge current at room temperature: Inverse spin-Hall effect. \emph{Applied physics letters} \textbf{2006}, \emph{88}\relax
\mciteBstWouldAddEndPuncttrue
\mciteSetBstMidEndSepPunct{\mcitedefaultmidpunct}
{\mcitedefaultendpunct}{\mcitedefaultseppunct}\relax
\EndOfBibitem
\bibitem[Tserkovnyak \latin{et~al.}(2002)Tserkovnyak, Brataas, and Bauer]{tserkovnyak2002spin}
Tserkovnyak,~Y.; Brataas,~A.; Bauer,~G.~E. Spin pumping and magnetization dynamics in metallic multilayers. \emph{Physical Review B} \textbf{2002}, \emph{66}, 224403\relax
\mciteBstWouldAddEndPuncttrue
\mciteSetBstMidEndSepPunct{\mcitedefaultmidpunct}
{\mcitedefaultendpunct}{\mcitedefaultseppunct}\relax
\EndOfBibitem
\bibitem[Shao \latin{et~al.}(2021)Shao, Li, Liu, Yang, Fukami, Razavi, Wu, Wang, Freimuth, Mokrousov, \latin{et~al.} others]{shao2021roadmap}
Shao,~Q.; Li,~P.; Liu,~L.; Yang,~H.; Fukami,~S.; Razavi,~A.; Wu,~H.; Wang,~K.; Freimuth,~F.; Mokrousov,~Y.; others Roadmap of spin--orbit torques. \emph{IEEE transactions on magnetics} \textbf{2021}, \emph{57}, 1--39\relax
\mciteBstWouldAddEndPuncttrue
\mciteSetBstMidEndSepPunct{\mcitedefaultmidpunct}
{\mcitedefaultendpunct}{\mcitedefaultseppunct}\relax
\EndOfBibitem
\bibitem[Roy \latin{et~al.}(2021)Roy, Mishra, Gupta, Mohanty, Singh, and Bedanta]{roy2021spin}
Roy,~K.; Mishra,~A.; Gupta,~P.; Mohanty,~S.; Singh,~B.~B.; Bedanta,~S. Spin pumping and inverse spin Hall effect in CoFeB/IrMn heterostructures. \emph{Journal of Physics D: Applied Physics} \textbf{2021}, \emph{54}, 425001\relax
\mciteBstWouldAddEndPuncttrue
\mciteSetBstMidEndSepPunct{\mcitedefaultmidpunct}
{\mcitedefaultendpunct}{\mcitedefaultseppunct}\relax
\EndOfBibitem
\bibitem[Singh \latin{et~al.}(2020)Singh, Roy, Chelvane, and Bedanta]{singh2020inverse}
Singh,~B.~B.; Roy,~K.; Chelvane,~J.~A.; Bedanta,~S. Inverse spin Hall effect and spin pumping in the polycrystalline noncollinear antiferromagnetic Mn 3 Ga. \emph{Physical Review B} \textbf{2020}, \emph{102}, 174444\relax
\mciteBstWouldAddEndPuncttrue
\mciteSetBstMidEndSepPunct{\mcitedefaultmidpunct}
{\mcitedefaultendpunct}{\mcitedefaultseppunct}\relax
\EndOfBibitem
\bibitem[Singh and Bedanta(2020)Singh, and Bedanta]{singh2020large}
Singh,~B.~B.; Bedanta,~S. Large spin Hall angle and spin-mixing conductance in the highly resistive antiferromagnet Mn 2 Au. \emph{Physical Review Applied} \textbf{2020}, \emph{13}, 044020\relax
\mciteBstWouldAddEndPuncttrue
\mciteSetBstMidEndSepPunct{\mcitedefaultmidpunct}
{\mcitedefaultendpunct}{\mcitedefaultseppunct}\relax
\EndOfBibitem
\bibitem[Kimata \latin{et~al.}(2019)Kimata, Chen, Kondou, Sugimoto, Muduli, Ikhlas, Omori, Tomita, MacDonald, Nakatsuji, \latin{et~al.} others]{kimata2019magnetic}
Kimata,~M.; Chen,~H.; Kondou,~K.; Sugimoto,~S.; Muduli,~P.~K.; Ikhlas,~M.; Omori,~Y.; Tomita,~T.; MacDonald,~A.~H.; Nakatsuji,~S.; others Magnetic and magnetic inverse spin Hall effects in a non-collinear antiferromagnet. \emph{Nature} \textbf{2019}, \emph{565}, 627--630\relax
\mciteBstWouldAddEndPuncttrue
\mciteSetBstMidEndSepPunct{\mcitedefaultmidpunct}
{\mcitedefaultendpunct}{\mcitedefaultseppunct}\relax
\EndOfBibitem
\bibitem[Singh \latin{et~al.}(2019)Singh, Jena, Samanta, Biswas, Satpati, and Bedanta]{singh2019inverse}
Singh,~B.~B.; Jena,~S.~K.; Samanta,~M.; Biswas,~K.; Satpati,~B.; Bedanta,~S. Inverse spin Hall effect in electron beam evaporated topological insulator Bi2Se3 thin film. \emph{physica status solidi (RRL)--Rapid Research Letters} \textbf{2019}, \emph{13}, 1800492\relax
\mciteBstWouldAddEndPuncttrue
\mciteSetBstMidEndSepPunct{\mcitedefaultmidpunct}
{\mcitedefaultendpunct}{\mcitedefaultseppunct}\relax
\EndOfBibitem
\bibitem[Jamali \latin{et~al.}(2015)Jamali, Lee, Jeong, Mahfouzi, Lv, Zhao, Nikolic, Mkhoyan, Samarth, and Wang]{jamali2015giant}
Jamali,~M.; Lee,~J.~S.; Jeong,~J.~S.; Mahfouzi,~F.; Lv,~Y.; Zhao,~Z.; Nikolic,~B.~K.; Mkhoyan,~K.~A.; Samarth,~N.; Wang,~J.-P. Giant spin pumping and inverse spin Hall effect in the presence of surface and bulk spin- orbit coupling of topological insulator Bi2Se3. \emph{Nano letters} \textbf{2015}, \emph{15}, 7126--7132\relax
\mciteBstWouldAddEndPuncttrue
\mciteSetBstMidEndSepPunct{\mcitedefaultmidpunct}
{\mcitedefaultendpunct}{\mcitedefaultseppunct}\relax
\EndOfBibitem
\bibitem[Chen and Yi(2021)Chen, and Yi]{chen2021spin}
Chen,~H.; Yi,~D. Spin--charge conversion in transition metal oxides. \emph{APL Materials} \textbf{2021}, \emph{9}\relax
\mciteBstWouldAddEndPuncttrue
\mciteSetBstMidEndSepPunct{\mcitedefaultmidpunct}
{\mcitedefaultendpunct}{\mcitedefaultseppunct}\relax
\EndOfBibitem
\bibitem[Sahoo \latin{et~al.}(2021)Sahoo, Roy, Gupta, Mishra, Satpati, Singh, and Bedanta]{sahoo2021spin}
Sahoo,~B.; Roy,~K.; Gupta,~P.; Mishra,~A.; Satpati,~B.; Singh,~B.~B.; Bedanta,~S. Spin pumping and inverse spin Hall effect in iridium oxide. \emph{Advanced Quantum Technologies} \textbf{2021}, \emph{4}, 2000146\relax
\mciteBstWouldAddEndPuncttrue
\mciteSetBstMidEndSepPunct{\mcitedefaultmidpunct}
{\mcitedefaultendpunct}{\mcitedefaultseppunct}\relax
\EndOfBibitem
\bibitem[Tokura and Nagaosa(2000)Tokura, and Nagaosa]{tokura2000orbital}
Tokura,~Y.; Nagaosa,~N. Orbital physics in transition-metal oxides. \emph{science} \textbf{2000}, \emph{288}, 462--468\relax
\mciteBstWouldAddEndPuncttrue
\mciteSetBstMidEndSepPunct{\mcitedefaultmidpunct}
{\mcitedefaultendpunct}{\mcitedefaultseppunct}\relax
\EndOfBibitem
\bibitem[Medvedeva and Hettiarachchi(2010)Medvedeva, and Hettiarachchi]{medvedeva2010tuning}
Medvedeva,~J.~E.; Hettiarachchi,~C.~L. Tuning the properties of complex transparent conducting oxides: Role of crystal symmetry, chemical composition, and carrier generation. \emph{Physical Review B—Condensed Matter and Materials Physics} \textbf{2010}, \emph{81}, 125116\relax
\mciteBstWouldAddEndPuncttrue
\mciteSetBstMidEndSepPunct{\mcitedefaultmidpunct}
{\mcitedefaultendpunct}{\mcitedefaultseppunct}\relax
\EndOfBibitem
\bibitem[Li \latin{et~al.}(2020)Li, Shi, Zhang, and MacManus-Driscoll]{li2020defects}
Li,~W.; Shi,~J.; Zhang,~K.~H.; MacManus-Driscoll,~J.~L. Defects in complex oxide thin films for electronics and energy applications: challenges and opportunities. \emph{Materials Horizons} \textbf{2020}, \emph{7}, 2832--2859\relax
\mciteBstWouldAddEndPuncttrue
\mciteSetBstMidEndSepPunct{\mcitedefaultmidpunct}
{\mcitedefaultendpunct}{\mcitedefaultseppunct}\relax
\EndOfBibitem
\bibitem[Cao and Wu(2011)Cao, and Wu]{cao2011strain}
Cao,~J.; Wu,~J. Strain effects in low-dimensional transition metal oxides. \emph{Materials Science and Engineering: R: Reports} \textbf{2011}, \emph{71}, 35--52\relax
\mciteBstWouldAddEndPuncttrue
\mciteSetBstMidEndSepPunct{\mcitedefaultmidpunct}
{\mcitedefaultendpunct}{\mcitedefaultseppunct}\relax
\EndOfBibitem
\bibitem[{\v{S}}mejkal \latin{et~al.}(2022){\v{S}}mejkal, Sinova, and Jungwirth]{vsmejkal2022emerging}
{\v{S}}mejkal,~L.; Sinova,~J.; Jungwirth,~T. Emerging research landscape of altermagnetism. \emph{Physical Review X} \textbf{2022}, \emph{12}, 040501\relax
\mciteBstWouldAddEndPuncttrue
\mciteSetBstMidEndSepPunct{\mcitedefaultmidpunct}
{\mcitedefaultendpunct}{\mcitedefaultseppunct}\relax
\EndOfBibitem
\bibitem[Feng \latin{et~al.}(2022)Feng, Zhou, {\v{S}}mejkal, Wu, Zhu, Guo, Gonz{\'a}lez-Hern{\'a}ndez, Wang, Yan, Qin, \latin{et~al.} others]{feng2022anomalous}
Feng,~Z.; Zhou,~X.; {\v{S}}mejkal,~L.; Wu,~L.; Zhu,~Z.; Guo,~H.; Gonz{\'a}lez-Hern{\'a}ndez,~R.; Wang,~X.; Yan,~H.; Qin,~P.; others An anomalous Hall effect in altermagnetic ruthenium dioxide. \emph{Nature Electronics} \textbf{2022}, \emph{5}, 735--743\relax
\mciteBstWouldAddEndPuncttrue
\mciteSetBstMidEndSepPunct{\mcitedefaultmidpunct}
{\mcitedefaultendpunct}{\mcitedefaultseppunct}\relax
\EndOfBibitem
\bibitem[Zhang \latin{et~al.}(2024)Zhang, Bai, Han, Chen, Zhou, Back, Pan, Wang, and Song]{zhang2024simultaneous}
Zhang,~Y.; Bai,~H.; Han,~L.; Chen,~C.; Zhou,~Y.; Back,~C.~H.; Pan,~F.; Wang,~Y.; Song,~C. Simultaneous High Charge-Spin Conversion Efficiency and Large Spin Diffusion Length in Altermagnetic RuO2. \emph{Advanced Functional Materials} \textbf{2024}, 2313332\relax
\mciteBstWouldAddEndPuncttrue
\mciteSetBstMidEndSepPunct{\mcitedefaultmidpunct}
{\mcitedefaultendpunct}{\mcitedefaultseppunct}\relax
\EndOfBibitem
\bibitem[Bose \latin{et~al.}(2022)Bose, Schreiber, Jain, Shao, Nair, Sun, Zhang, Muller, Tsymbal, Schlom, \latin{et~al.} others]{bose2022tilted}
Bose,~A.; Schreiber,~N.~J.; Jain,~R.; Shao,~D.-F.; Nair,~H.~P.; Sun,~J.; Zhang,~X.~S.; Muller,~D.~A.; Tsymbal,~E.~Y.; Schlom,~D.~G.; others Tilted spin current generated by the collinear antiferromagnet ruthenium dioxide. \emph{Nature Electronics} \textbf{2022}, \emph{5}, 267--274\relax
\mciteBstWouldAddEndPuncttrue
\mciteSetBstMidEndSepPunct{\mcitedefaultmidpunct}
{\mcitedefaultendpunct}{\mcitedefaultseppunct}\relax
\EndOfBibitem
\bibitem[Bai \latin{et~al.}(2022)Bai, Han, Feng, Zhou, Su, Wang, Liao, Zhu, Chen, Pan, \latin{et~al.} others]{bai2022observation}
Bai,~H.; Han,~L.; Feng,~X.; Zhou,~Y.; Su,~R.; Wang,~Q.; Liao,~L.; Zhu,~W.; Chen,~X.; Pan,~F.; others Observation of spin splitting torque in a collinear antiferromagnet RuO 2. \emph{Physical Review Letters} \textbf{2022}, \emph{128}, 197202\relax
\mciteBstWouldAddEndPuncttrue
\mciteSetBstMidEndSepPunct{\mcitedefaultmidpunct}
{\mcitedefaultendpunct}{\mcitedefaultseppunct}\relax
\EndOfBibitem
\bibitem[Kittel(1948)]{kittel1948theory}
Kittel,~C. On the theory of ferromagnetic resonance absorption. \emph{Physical review} \textbf{1948}, \emph{73}, 155\relax
\mciteBstWouldAddEndPuncttrue
\mciteSetBstMidEndSepPunct{\mcitedefaultmidpunct}
{\mcitedefaultendpunct}{\mcitedefaultseppunct}\relax
\EndOfBibitem
\bibitem[Brataas \latin{et~al.}(2002)Brataas, Tserkovnyak, Bauer, and Halperin]{brataas2002spin}
Brataas,~A.; Tserkovnyak,~Y.; Bauer,~G.~E.; Halperin,~B.~I. Spin battery operated by ferromagnetic resonance. \emph{Physical Review B} \textbf{2002}, \emph{66}, 060404\relax
\mciteBstWouldAddEndPuncttrue
\mciteSetBstMidEndSepPunct{\mcitedefaultmidpunct}
{\mcitedefaultendpunct}{\mcitedefaultseppunct}\relax
\EndOfBibitem
\bibitem[Conca \latin{et~al.}(2017)Conca, Heinz, Schweizer, Keller, Papaioannou, and Hillebrands]{conca2017lack}
Conca,~A.; Heinz,~B.; Schweizer,~M.; Keller,~S.; Papaioannou,~E.~T.; Hillebrands,~B. Lack of correlation between the spin-mixing conductance and the inverse spin Hall effect generated voltages in CoFeB/Pt and CoFeB/Ta bilayers. \emph{Physical Review B} \textbf{2017}, \emph{95}, 174426\relax
\mciteBstWouldAddEndPuncttrue
\mciteSetBstMidEndSepPunct{\mcitedefaultmidpunct}
{\mcitedefaultendpunct}{\mcitedefaultseppunct}\relax
\EndOfBibitem
\bibitem[Mosendz \latin{et~al.}(2010)Mosendz, Vlaminck, Pearson, Fradin, Bauer, Bader, and Hoffmann]{mosendz2010detection}
Mosendz,~O.; Vlaminck,~V.; Pearson,~J.; Fradin,~F.; Bauer,~G.; Bader,~S.; Hoffmann,~A. Detection and quantification of inverse spin Hall effect from spin pumping in permalloy/normal metal bilayers. \emph{Physical Review B—Condensed Matter and Materials Physics} \textbf{2010}, \emph{82}, 214403\relax
\mciteBstWouldAddEndPuncttrue
\mciteSetBstMidEndSepPunct{\mcitedefaultmidpunct}
{\mcitedefaultendpunct}{\mcitedefaultseppunct}\relax
\EndOfBibitem
\bibitem[Vlaminck \latin{et~al.}(2013)Vlaminck, Pearson, Bader, and Hoffmann]{vlaminck2013dependence}
Vlaminck,~V.; Pearson,~J.~E.; Bader,~S.~D.; Hoffmann,~A. Dependence of spin-pumping spin Hall effect measurements on layer thicknesses and stacking order. \emph{Physical Review B—Condensed Matter and Materials Physics} \textbf{2013}, \emph{88}, 064414\relax
\mciteBstWouldAddEndPuncttrue
\mciteSetBstMidEndSepPunct{\mcitedefaultmidpunct}
{\mcitedefaultendpunct}{\mcitedefaultseppunct}\relax
\EndOfBibitem
\bibitem[Shaw \latin{et~al.}(2012)Shaw, Nembach, and Silva]{shaw2012determination}
Shaw,~J.~M.; Nembach,~H.~T.; Silva,~T.~J. Determination of spin pumping as a source of linewidth in sputtered Co 90 Fe 10/Pd multilayers by use of broadband ferromagnetic resonance spectroscopy. \emph{Physical Review B—Condensed Matter and Materials Physics} \textbf{2012}, \emph{85}, 054412\relax
\mciteBstWouldAddEndPuncttrue
\mciteSetBstMidEndSepPunct{\mcitedefaultmidpunct}
{\mcitedefaultendpunct}{\mcitedefaultseppunct}\relax
\EndOfBibitem
\bibitem[Foros \latin{et~al.}(2005)Foros, Woltersdorf, Heinrich, and Brataas]{foros2005scattering}
Foros,~J.; Woltersdorf,~G.; Heinrich,~B.; Brataas,~A. Scattering of spin current injected in Pd (001). \emph{Journal of applied physics} \textbf{2005}, \emph{97}\relax
\mciteBstWouldAddEndPuncttrue
\mciteSetBstMidEndSepPunct{\mcitedefaultmidpunct}
{\mcitedefaultendpunct}{\mcitedefaultseppunct}\relax
\EndOfBibitem
\bibitem[Lee and Yang(2011)Lee, and Yang]{lee2011progress}
Lee,~J.-K.; Yang,~M. Progress in light harvesting and charge injection of dye-sensitized solar cells. \emph{Materials Science and Engineering: B} \textbf{2011}, \emph{176}, 1142--1160\relax
\mciteBstWouldAddEndPuncttrue
\mciteSetBstMidEndSepPunct{\mcitedefaultmidpunct}
{\mcitedefaultendpunct}{\mcitedefaultseppunct}\relax
\EndOfBibitem
\bibitem[Rojas-S{\'a}nchez \latin{et~al.}(2014)Rojas-S{\'a}nchez, Reyren, Laczkowski, Savero, Attan{\'e}, Deranlot, Jamet, George, Vila, and Jaffr{\`e}s]{rojas2014spin}
Rojas-S{\'a}nchez,~J.-C.; Reyren,~N.; Laczkowski,~P.; Savero,~W.; Attan{\'e},~J.-P.; Deranlot,~C.; Jamet,~M.; George,~J.-M.; Vila,~L.; Jaffr{\`e}s,~H. Spin pumping and inverse spin Hall effect in platinum: the essential role of spin-memory loss at metallic interfaces. \emph{Physical review letters} \textbf{2014}, \emph{112}, 106602\relax
\mciteBstWouldAddEndPuncttrue
\mciteSetBstMidEndSepPunct{\mcitedefaultmidpunct}
{\mcitedefaultendpunct}{\mcitedefaultseppunct}\relax
\EndOfBibitem
\bibitem[Khasawneh \latin{et~al.}(2011)Khasawneh, Klose, Pratt~Jr, and Birge]{khasawneh2011spin}
Khasawneh,~M.~A.; Klose,~C.; Pratt~Jr,~W.; Birge,~N.~O. Spin-memory loss at Co/Ru interfaces. \emph{Physical Review B—Condensed Matter and Materials Physics} \textbf{2011}, \emph{84}, 014425\relax
\mciteBstWouldAddEndPuncttrue
\mciteSetBstMidEndSepPunct{\mcitedefaultmidpunct}
{\mcitedefaultendpunct}{\mcitedefaultseppunct}\relax
\EndOfBibitem
\bibitem[Singh \latin{et~al.}(2021)Singh, Roy, Gupta, Seki, Takanashi, and Bedanta]{singh2021high}
Singh,~B.~B.; Roy,~K.; Gupta,~P.; Seki,~T.; Takanashi,~K.; Bedanta,~S. High spin mixing conductance and spin interface transparency at the interface of a Co2Fe0. 4Mn0. 6Si Heusler alloy and Pt. \emph{NPG Asia Materials} \textbf{2021}, \emph{13}, 9\relax
\mciteBstWouldAddEndPuncttrue
\mciteSetBstMidEndSepPunct{\mcitedefaultmidpunct}
{\mcitedefaultendpunct}{\mcitedefaultseppunct}\relax
\EndOfBibitem
\bibitem[Panda \latin{et~al.}(2021)Panda, Majumder, Bhattacharyya, Dutta, Choudhury, and Barman]{panda2021structural}
Panda,~S.~N.; Majumder,~S.; Bhattacharyya,~A.; Dutta,~S.; Choudhury,~S.; Barman,~A. Structural phase-dependent giant interfacial spin transparency in W/CoFeB thin-film heterostructures. \emph{ACS Applied Materials \& Interfaces} \textbf{2021}, \emph{13}, 20875--20884\relax
\mciteBstWouldAddEndPuncttrue
\mciteSetBstMidEndSepPunct{\mcitedefaultmidpunct}
{\mcitedefaultendpunct}{\mcitedefaultseppunct}\relax
\EndOfBibitem
\bibitem[Zhang \latin{et~al.}(2015)Zhang, Han, Jiang, Yang, and SP~Parkin]{zhang2015role}
Zhang,~W.; Han,~W.; Jiang,~X.; Yang,~S.-H.; SP~Parkin,~S. Role of transparency of platinum--ferromagnet interfaces in determining the intrinsic magnitude of the spin Hall effect. \emph{Nature physics} \textbf{2015}, \emph{11}, 496--502\relax
\mciteBstWouldAddEndPuncttrue
\mciteSetBstMidEndSepPunct{\mcitedefaultmidpunct}
{\mcitedefaultendpunct}{\mcitedefaultseppunct}\relax
\EndOfBibitem
\end{mcitethebibliography}

\end{document}